\documentclass{rstransa} 
\usepackage{amssymb,amsmath}
\usepackage{graphicx}

\begin{document}

\title{Two-dimensional turbulent condensates without bottom drag}

\author{
Adrian van Kan$^{1}$,
Alexandros Alexakis$^{2}$ and Edgar Knobloch$^{1}$}

\address{$^{1}$ Department of Physics, University of California at Berkeley, Berkeley, California 94720, USA\\
$^{2}$ Laboratoire de Physique de l’Ecole Normale Sup\'e erieure, ENS, Universit\'e PSL, CNRS,
 Sorbonne Universit\'e, Universit\'e Paris-Diderot, Sorbonne Paris Cit\'e, Paris, France}

\subject{fluid mechanics, statistical physics}


\corres{Adrian van Kan\\
\email{avankan@berkeley.edu}}

\begin{abstract}\sloppypar
The extent to which statistical equilibrium theory is applicable to driven dissipative dynamics remains an important open question in many systems. We use extensive direct numerical simulations of the incompressible two-dimensional (2D) Navier-Stokes equation to examine the steady state of large-scale condensates in 2D turbulence at finite Reynolds number $Re$ in the absence of bottom drag.
Large-scale condensates appear above a critical Reynolds number $Re_c\approx 4.19$. 
Close to this onset, we find 
{a power-law}
scaling of the 
energy with $Re-Re_c$, with the energy spectrum at large scales following the absolute equilibrium form proposed by Kraichnan. At larger $Re$, the energy spectrum deviates from this 
form, displaying a steep power-law range at low wave numbers with exponent $-5$, with most of the energy dissipation occurring within the condensate at large scales. We show that this spectral exponent is consistent with the logarithmic radial vorticity profile of the condensate vortices predicted by quasi-linear theory for a viscously saturated condensate. Our findings shed new light on the classical problem of large-scale turbulent condensation in forced dissipative 2D flows in finite domains, showing that the large scales are close to equilibrium dynamics in weakly turbulent flows but not in the strong condensate regime with $Re\gg1$.
\end{abstract}
\maketitle

\section{Introduction}
Turbulence is a highly chaotic state of fluid flow at large Reynolds numbers \cite{frisch1995turbulence} that involves many degrees of freedom and therefore requires a statistical description. In this approach mean field quantities of interest should be derivable without detailed knowledge of the microscopic dynamics, much as the equation of state for a macroscopic system may be derived without detailed knowledge of individual molecular motions. 
However, approaches adapted from classical equilibrium statistical physics fail in general when applied to turbulence because
the turbulent flow is typically far from thermodynamic equilibrium \cite{ruelle2012hydrodynamic,goldenfeld2017turbulence}, with nonzero energy fluxes across scales and finite dissipation rates in the zero viscosity limit, both of which break time reversibility.
Nevertheless, there are instances where the mean fluxes are negligible compared to fluctuations, rendering turbulence close to an equilibrium state, at least over some range of scales. In such cases approaches from equilibrium statistical physics may be pursued.

For three-dimensional turbulence,
Lee \cite{lee1952some} was the first to apply equilibrium statistical physics to flows described by an ensemble of Fourier modes, arriving at a Rayleigh-Jeans spectrum $E(k)\propto k^2$. This work was later extended by Kraichnan, who incorporated the helicity constraint \cite{kraichnan1973helical}. Lee's and Kraichnan's results strictly apply to what
is now known as the {\it truncated Euler equations} \cite{cichowlas2005effective} where a discrete set of wave numbers with finite modulus is retained in the evolution governed by the Euler equation of ideal (i.e., unforced and dissipationless) fluid flow, leading to a finite-dimensional dynamical system. However, these results apply in other instances as well. First, it has been argued theoretically \cite{forster1977large} and demonstrated both numerically and experimentally \cite{dallas2015statistical,cameron2017effect,alexakis2019thermal,gorce2022statistical,alexakis2023fluctuation}
that the Rayleigh-Jeans spectrum provides an adequate description of the scales above the forcing scale, with deviations from this spectrum studied in
\cite{hosking2023emergence,ding2024departure}.
Second, at the smallest scales of three-dimensional turbulence, it has been argued \cite{frisch2008hyperviscosity} and
recently shown numerically \cite{agrawal2020turbulent} that when a high-order hyperviscosity is used the energy spectrum just above the hyperviscous scales forms a $k^2$ spectrum. Finally, when the Navier-Stokes equation is solved
spectrally on a periodic domain in three dimensions, retaining a finite number of Fourier modes, it was shown
in \cite{alexakis2020energy} that, in the limit of small viscosity,
the sharp spectral cutoff also leads to a $k^2$ spectrum, much like the effect of high-order hyperviscosity.
In the latter two cases, the $k^2$ range forms due to an artificial block, also referred to as a bottleneck. 
This artificial block forces the energy flux to fall abruptly to zero at the largest available wave numbers 
leading the flow to take a near-equilibrium functional form. 

In two-dimensional turbulence, where energy is transferred towards larger scales \cite{boffetta2012two}, there is a natural, and physically significant, smallest wave number in the system set by the inverse of the domain size. In the absence of an efficient large-scale dissipation mechanism, such as bottom drag, energy piles up at the largest scales, producing large-scale flow structures, also known as condensates \cite{sommeria1986experimental,smith1993bose,falkovich1992inverse}. The weak dissipation and the reduced energy flux due to the low spectral cutoff both suggest that these scales might be described by equilibrium dynamics.
The approach to describing such large-scale organisation in two-dimensional fluid flows using equilibrium statistical physics goes back to
Onsager \cite{onsager1949statistical, eyink2006onsager} who was the first to analyze two-dimensional, discrete point-vortex flows in terms of microcanonical statistical mechanics, revealing the existence of states with negative temperature, where point vortices of either sign cluster to form large-scale vortices. This work was further pursued, in the specific context of point vortex dynamics in \cite{lundgren1977statistical,chavanis1996statistical} and later generalised to continuous vorticity fields obeying the two-dimensional Euler equation in the so-called Robert-Sommeria-Miller theory \cite{robert1991statistical,miller1992statistical}, which has been studied intensively \cite{bouchet2012statistical} with the goal of predicting the long time evolution of turbulent large-scale structures.

In an alternative approach built on the truncated Euler equation, inspired by Lee's work in three dimensions, Kraichnan \cite{kraichnan1967inertial,kraichnan1975statistical} formulated an absolute equilibrium theory for two-dimensional turbulence. 
He assumed that the probability of a given state follows a generalised Gibbs distribution familiar from canonical statistical mechanics, taking into account an additional ideal quadratic invariant, the enstrophy $\Omega$ (mean squared vorticity), together with the energy $\mathcal{E}$. This led to a predicted energy spectrum of the form 
\begin{equation}
    E(k)= \frac{2\pi k}{\alpha + \beta k^2},\label{eq:kraichnan_spectrum2D}
\end{equation}
where the constants $\alpha$ and $\beta$ are generalised temperatures. These may be positive or negative, and their values are determined by the total energy and enstrophy given by 
\begin{equation}
    {\mathcal{E}} = \sum_\mathbf{k} E(k), \hspace{2cm} \Omega= \sum_\mathbf{k} k^2 E(k), \label{eq:en_ens_spec}
\end{equation}
where the sum is over all wave numbers in the range  $k_{min} \le |{\bf k}| \le k_{max}$,
with the minimum wave number $k_{min}$ (largest length scale) and the maximum wave number $k_{max}$ (smallest length scale). 
In addition to the canonical ensemble adopted by Kraichnan, the statistical mechanics of the two-dimensional truncated Euler equation has also been investigated in the microcanonical ensemble to predict, for instance, reversal statistics of a confined large-scale vortex \cite{shukla2016statistical,van2022geometric}. The coexistence of a Kraichnan-type absolute equilibrium at small scales and a point-vortex type equilibrium at large scales was recently reported in \cite{agoua2025coexistence}.

While Kraichnan's absolute equilibrium theory assumes that the statistics of the flow are well described by the truncated Euler equations of ideal (unforced and dissipationless) fluid flow, it remains ill-understood how this relates to forced-dissipative turbulence. Early numerical studies investigated primarily the dynamics of freely decaying two-dimensional turbulence \cite{lilly1971numerical,brachet1988dynamics,weiss1993temporal,bracco2000revisiting}, in part due to limited available computational resources. Other work focused on the cascade phenomenology 
{of forced two-dimensional turbulence 
by introducing a sufficiently strong bottom drag (also referred to as Rayleigh friction) that prevented the formation of }
a large-scale condensate, e.g. \cite{boffetta2007energy}. The focus of many of the more recent numerical simulations, on the other hand, was on the forced-dissipative condensate state \cite{smith1993bose, boffetta2012two,chan2012dynamics,frishman2018turbulence,laurie2014universal,frishman2017jets,xu2024fluctuation,parfenyev2024statistical}, where a well-defined statistically steady state is attained at late times. Chertkov et al.~\cite{chertkov2007dynamics} studied the transient dynamics of the build-up of the condensate under sustained forcing, by integrating the two-dimensional Navier-Stokes equation subject to periodic boundary conditions with a small-scale stochastic forcing, hyperviscosity and no large-scale bottom drag, producing (by construction) a transient simulation where energy kept increasing. Chertkov et al. reported a $k^{-3}$ power-law range in $E(k)$ at small $k$ associated with the large-scale condensate (a vortex dipole in physical space). However, the limitation of these results to the transient regime motivates the question of how this result is modified in a statistically stationary state where the condensate amplitude saturates due to dissipation at large scales. Such steady-state condensate states have been studied in recent years in the context of quasi-linear theory \cite{laurie2014universal,frishman2017culmination,kolokolov2020coherent,parfenyev2022profile,svirsky2023statistics}, taking advantage of the strength of the mean flow compared to fluctuations to neglect nonlinear interactions between fluctuations except where they feed back directly on the mean flow. This theoretical work has led to explicit predictions for the profile of the large-scale condensate, with results that depend on the dissipation mechanism acting at the large scales. When bottom drag saturates the condensate there is, at large Reynolds numbers, a range of radii with constant mean azimuthal velocity  \cite{laurie2014universal}, while in the case of a viscously saturated condensate
one finds a mean azimuthal velocity profile of the form $r\ln(r/R)$ \cite{doludenko2021coherent}, where $r$ is the radial coordinate and $R$ is the size of the condensate and comparable to the domain size. These predictions about the real-space profile of the condensate have been found to compare favorably with direct numerical simulations (DNS) when the condensate is strong. 

In view of the long history of statistical mechanics-based approaches to describing the condensate in the context of an ideal fluid, an important open question remains, namely, whether the large scales in forced-dissipative two-dimensional turbulent condensates can in fact be described by equilibrium dynamics and, if so, under which conditions? Below, we present a detailed exploration of this question, focusing on the case without bottom drag with condensates of sufficiently large amplitude that their growth saturates viscously in the statistically steady state. Our results indicate that close to the onset of two-dimensional turbulence, where the upscale energy flux is small, the large scales do indeed display a Kraichnan equilibrium form in well-resolved DNS but only up to some finite Reynolds number $Re$. For larger values of $Re$, the condensate remains far from equilibrium and quasi-linear theory applies, which is shown to imply an energy spectrum with a power-law form close to $k^{-5}$ at low wave numbers $k$.

 The remainder of this article is structured as follows. In Sec.~\ref{sec:setup} we describe our setup and discuss the relevant control parameters. In Sec.~\ref{sec:2D_NSE}, we present our numerical results, including  steady-state kinetic energy, energy and enstrophy dissipation rates, energy spectra and fluxes, as well as the relation of the observed energy spectra to the logarithmic radial physical space vorticity profile found in the viscously saturated condensate. Finally, in Sec.~\ref{sec:conclusions} we discuss the significance of our results and perspectives for future work.

\section{Set-up\label{sec:setup}} 

We consider incompressible flow in a two-dimensional, doubly periodic square domain of side length $2\pi L$. The flow obeys the incompressible two-dimensional vorticity equation 
\begin{equation}
    \partial_t \omega + {\bf u} \cdot {\bf \nabla } \omega = \nu \nabla^2 \omega + f_\omega , \hspace{1cm} \nabla \cdot \mathbf{u} = 0,
    \label{eq:NS}
\end{equation}
where ${\bf u}\equiv (u_x,u_y)$ is the velocity field, $\omega\equiv\partial_x u_y - \partial_y u_x$ is the vertical component of vorticity, $\nu$ is the fluid's kinematic viscosity and $f_\omega$ is the vertical component of the curl of an external 
body force. The forcing $f_\omega$ is composed of Fourier modes with wave vectors $\bf k$ 
lying in the isotropic ring $k_f \le |{\bf k}| \le k_f+\Delta k_f$  (where $\Delta k_f L=4$), whose amplitudes evolve
randomly, being $\delta$-correlated in time, so that the energy injection rate is fixed to $\varepsilon$,
while the enstrophy injection rate $\eta$ is fixed to $\eta\simeq \varepsilon k_f^2$ 
(the equality here becomes exact when $\Delta k_f =0$). 
The above system is characterised by only two non-dimensional numbers, namely,
the Reynolds number {based on the energy injection rate $\varepsilon$ and the forcing scale $k_f$}, viz. $Re\equiv \varepsilon^{1/3} /(k_f^{4/3}\nu)$,
%
and the scale separation factor $\Lambda\equiv k_fL$ between the forcing scale and the domain size.

The incompressible vorticity equation is solved using the pseudo-spectral code {\sc Ghost} \cite{mininni2011hybrid}
with a 2/3 de-aliasing rule and a second-order Runge-Kutta time advancement method on an evenly spaced grid of size $N$ in each direction. The finite resolution introduces a third non-dimensional number $k_{max}\eta$, 
where $k_{max}=N/(3L)$ is the maximum allowed wave number and $\eta \equiv \varepsilon^{-1/6}k_f^{-1/3}\nu^{1/2}$
is the enstrophy dissipation scale. For well-resolved runs $k_{max}\eta\gg1$, which we ensure in all simulations.
In the above set-up the smallest wave number $k_{min}$ is such that $k_{min}L=1$. The second smallest wave number is such that $|{\bf k}|L=\sqrt{2}$ which is significantly larger than $k_{min}$ and may hamper comparison with predictions for the functional form of the spectrum in cases where a strong condensate forms such that $k_{min}^2 \gtrsim \alpha/\beta$. For this reason we also consider cases for which $k_{min}L>1$, a situation achieved by artificially removing all wave numbers with $|{\bf k}|L < k_{min}L$ where $k_{min}\ge 1/L$ is chosen arbitrarily. 
This procedure allows us to reduce the spacing between the smallest and second smallest wave numbers, 
leading to a nearly continuous distribution of wave numbers and introducing a fourth 
non-dimensional number $\lambda=k_{min}L\ge1$.


{A large set of simulations was produced varying all four parameters.
The scale separation $\Lambda$ was varied between $\Lambda=2$ and $\Lambda=80$ while $\lambda$ varied from $\lambda=1$ to $\lambda=10$. The Reynolds number values achieved ranged from $Re\simeq1$ to $Re\simeq 60$. The majority of the runs employed a resolution $N=1024$ with $N=512$ grid points used only for $Re<4$ while grids with $N=2048$ were used for the two highest values of $Re$ examined in the $\Lambda\le 40$ runs. The slow convergence in time to the statistically stationary state prevented us from examining larger values of $Re$ with larger values of $N$. }
All runs were integrated until a statistically stationary state was reached, with the injection of kinetic energy balanced by viscous dissipation and all quantities fluctuating around a well-defined mean value. 

\section{Results \label{sec:2D_NSE}}  
\begin{figure}
    \centering
    \includegraphics[width=1.05\linewidth]{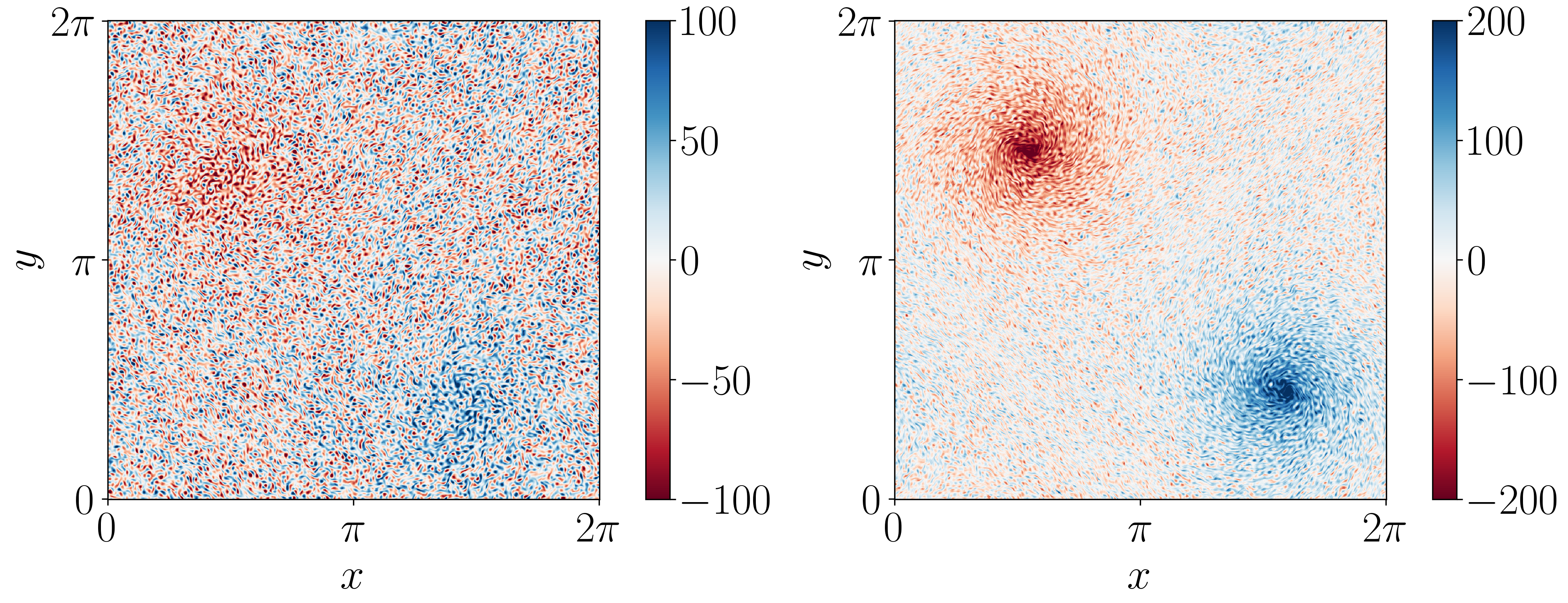}
    \caption{Vorticity field in the large-scale condensate in the statistically steady state with $\Lambda=80$, $k_{min}L=1$ and 
 $Re=4.32$ (left panel) and $Re=7.25$ (right panel). Close to the onset of turbulence (left panel), the condensate is weak and diffuse, and barely visible relative to the sea of smaller-scale vortices. At larger Reynolds numbers (right panel) the condensate becomes sharply peaked within the cores of the counter-rotating large-scale vortices while the smaller-scale vortices in the interstitial space between them remain but are subdominant. }
    \label{fig:ill}
\end{figure}

Here, we describe the results of our direct numerical simulations in terms of steady-state energy, energy and enstrophy dissipation rates, spectra and fluxes, as well as the spatial profile of the vortices comprising the large-scale condensate. The physical-space profile of the condensate is illustrated in Fig.~\ref{fig:ill} in terms of the vorticity field at two different 
Reynolds numbers $Re=4.32$ (left panel, close to the onset of turbulence) and $Re=7.25$ (right panel, strongly turbulent flow). 
While the condensate at $Re=4.32$ is weak and diffuse, and barely visible in the vorticity field relative to the background of smaller-scale vortices, at $Re=7.25$ a large-scale, large-amplitude vorticity dipole has clearly formed over the sea of smaller-scale vortices, which remain visible between the large-scale dipole. The purpose of this work is to describe this condensate state as a function of the parameters $Re$ and $\Lambda$.

\subsection{Global behaviour}\label{sec:Global}

\begin{figure}                                                                
  \begin{center}
      (a) \hspace{5cm} (b) 
\end{center}
\centerline{\includegraphics[width=0.50\textwidth]{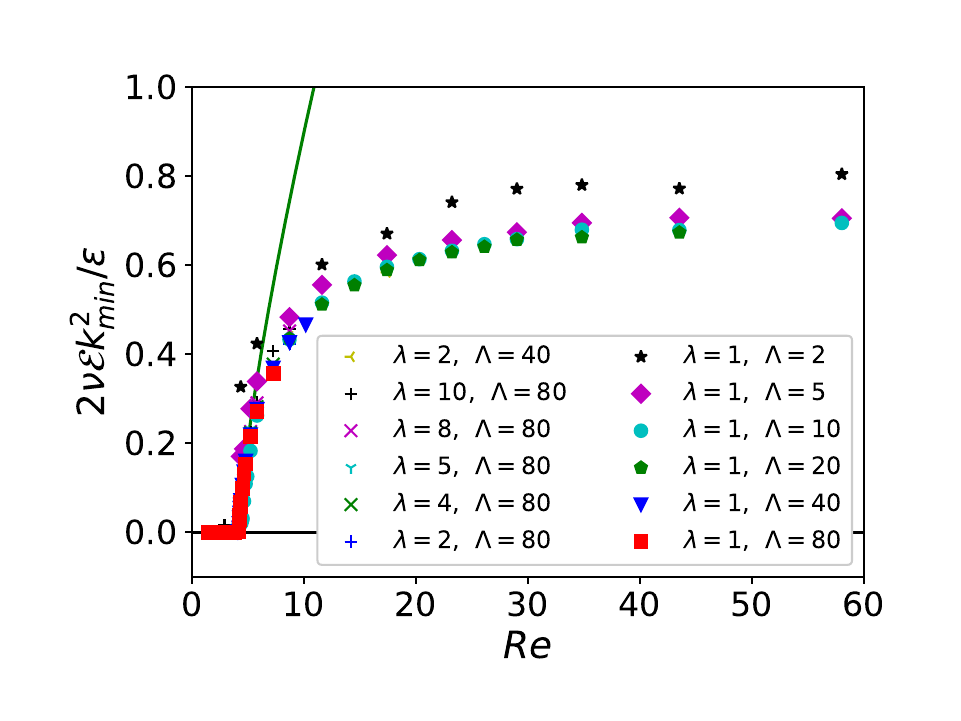}
              \includegraphics[width=0.50\textwidth]{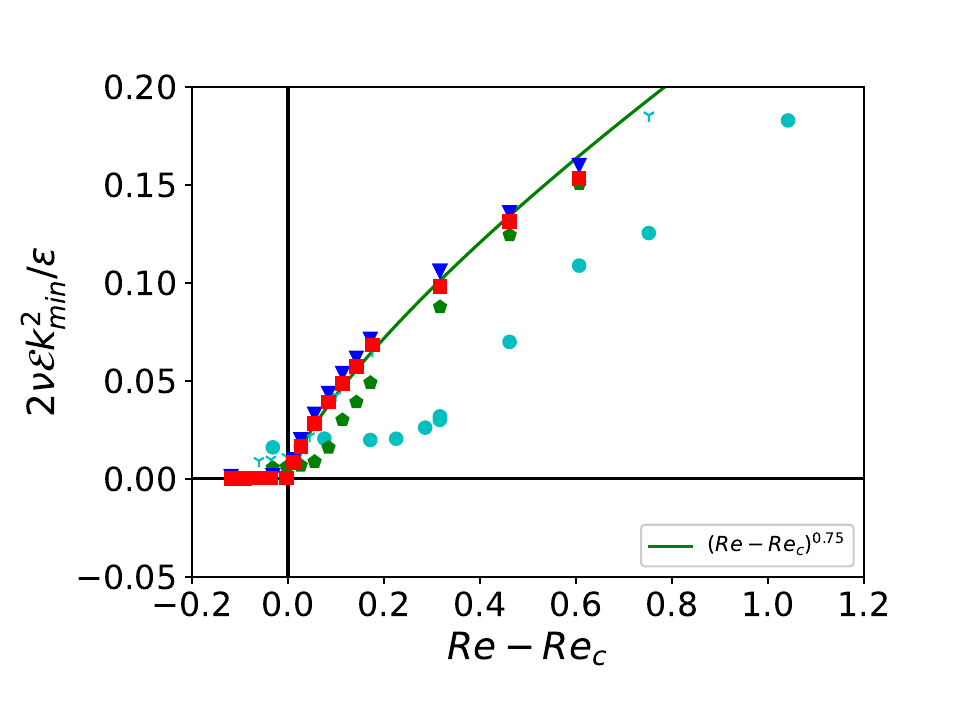} }
  \caption{Energy bifurcation diagram. Left: Energy normalised by $\varepsilon/\nu k_{min}^2$ over the range of $Re$ examined for different values of the parameters $\lambda$ and $\Lambda$.
  Right: Zoom of the onset region for turbulence. The green line shows a fit to the $\lambda\equiv k_{min}L=1$, $\Lambda\equiv k_f L=80$ data. }
\label{fig:Energy}
\end{figure}

The left panel of Fig.~\ref{fig:Energy} shows the total kinetic energy $\mathcal{E}=\frac{1}{2}\langle |\mathbf{u}|^2 \rangle$ of the flow (where $\langle \cdot \rangle$ is the average over time and space) as a function of $Re$ for all values of $\Lambda$ {and $\lambda$} examined. 
Energy is normalised by $\varepsilon / (2 \nu k_{min}^2)$ which collapses the data  provided there is sufficient scale separation between the forcing scale and the largest scale in the domain, i.e., $\Lambda\gg1$. The figure shows that the only cases to escape this data collapse are those with small values of the scale separation factor $\Lambda$, viz. $\Lambda=5$ and $\Lambda=2$.

%
The behaviour of the energy as a function of $Re$ constitutes a bifurcation diagram.
For $Re$ smaller than a
critical value $Re_c\simeq 4.19$ the normalised energy assumes very small (albeit finite) values, while for $Re>Re_c$ the kinetic energy bifurcates towards order one magnitudes.  
The right panel of Fig.~\ref{fig:Energy} shows a zoom of the same data focusing on the vicinity of $Re=Re_c$. 
The critical value $Re_c$ depends weakly on the scale separation factor $\Lambda$ but converges to a $\Lambda$-independent value as $\Lambda\to\infty$.
This behaviour is consistent with a transition from positive to negative eddy viscosity $\nu_e(k)=\beta(Re)+\delta\, k^2 + \dots$ \cite{dubrulle1991eddy}, where the first term $\beta(Re)$ changes sign at $Re_c$. 
For $Re<Re_c$ the bifurcation is imperfect and
the normalised energy is not zero, but as the scale separation $\Lambda$ increases, it becomes smaller and smaller approaching a perfect bifurcation. For $Re>Re_c$, the energy amplitude follows a $(Re-Re_c)^\gamma$ power-law behaviour, where the exponent $\gamma$ appears to be strictly smaller than
one, with $\gamma\approx0.75$. This exponent differs from that reported by Linkmann et al. \cite{linkmann2020non}, where $\gamma$ was found to approach unity as the bottom drag is reduced. However, it has to be stressed that computing critical exponents by a fitting procedure involves significant uncertainty; moreover, the forcing function considered by Linkmann et al. acts over a broader band of wave numbers than that considered here. We mention that for a one-dimensional deterministic bifurcation, one would expect $\gamma=1$,
but in the present case the presence of noise as well as the fact that the large-scale energy is distributed among a set of wave numbers and not a single mode may combine to lead to a different exponent. 

{ We note that the energy satisfies the inequality $2 \nu \mathcal{E}k_{min}^2/\varepsilon \le 1$, a consequence of Poincaré's inequality 
$\langle |\nabla {\bf u}|^2\rangle \ge k_{min}^2 \langle |{\bf u}|^2 \rangle$. This bound is sharp when all energy is concentrated in the $k=k_{min}$ modes. }
Our simulations show that this upper bound is missed by a substantial margin even at large values of $Re$.
This fact indicates that {the large scale energy in steady state is proportional to $1/\nu$ and, moreover, that
the smaller scales continue to contain a finite energy fraction even in the limit $Re\to\infty$.}  

\begin{figure}                                                                   
  \centerline{\includegraphics[width=0.50\textwidth]{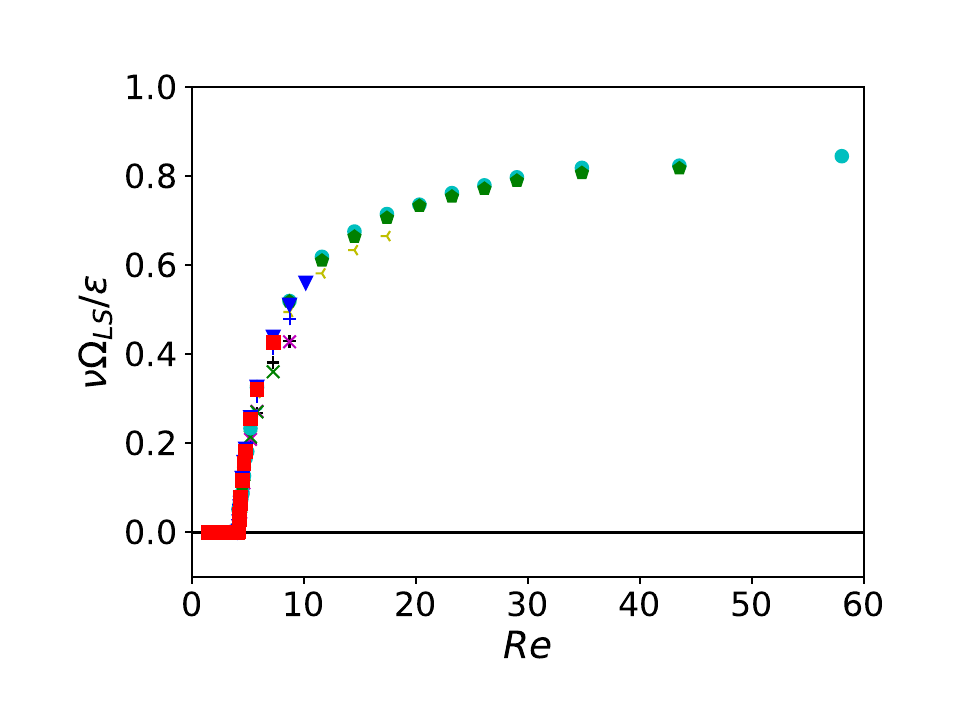}
              \includegraphics[width=0.50\textwidth]{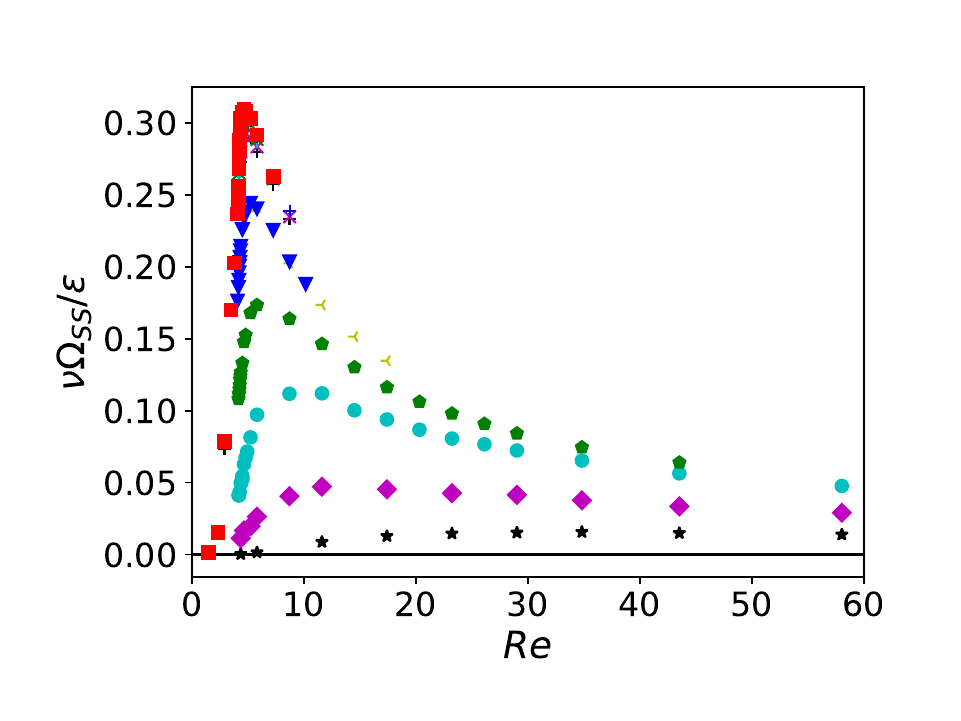} }
  \caption{Large-scale and small-scale enstrophy bifurcation diagrams as a function of $Re$, normalized by $\varepsilon/\nu$. Left: Large-scale enstrophy $\Omega_{LS}$ defined as the enstrophy in modes with $|k|\le k_{min}+3/L$. Right: Small-scale enstrophy $\Omega_{SS}$ defined as the enstrophy in modes with $|k|\ge k_f+6/L$.}
\label{fig:Enstrophy}
\end{figure}

The steady state of the system follows the energy and enstrophy balance relations 
\begin{equation}
    \varepsilon = \nu \Omega \quad \mathrm{and} \quad \eta=\nu \mathcal{Z} \simeq \varepsilon k_f^2 ,
    \label{OZ}
 \end{equation} 
where $\Omega=\langle |\nabla {\bf u}|^2\rangle$ is the enstrophy and $\mathcal{Z}=\langle  |\nabla \omega |^2\rangle$ is the palinstrophy.
These two relations constrain the flow such that most energy is transferred to large scales while most enstrophy is transferred to small scales \cite{eyink1996exact,alexakis2006energy}.
In Fig.~\ref{fig:Enstrophy} we plot the normalised energy dissipation $\nu \Omega_{LS}/\varepsilon$  
restricted to the smallest wave numbers $|{\bf k}|\le k_{min}+3/L$ while in the right panel we plot the normalised energy dissipation $\nu \Omega_{SS}/\varepsilon$ restricted to large wave numbers such that $|{\bf k}|\ge k_f+6/L$. As with the energy, the large-scale energy dissipation $\nu\Omega_{LS}$ 
increases above $Re_c$ and approaches a value close to (but smaller than) one at large $Re$.
In contrast, the small-scale dissipation initially increases but then falls to near zero values with increasing $Re$. 
Thus at large $Re$, most of the energy is dissipated at the largest scales. 
Note that this is true despite the fact that no large scale drag has been used
and all dissipation is due to viscosity.
This is achieved by the condensate reaching extremely high velocity amplitudes at small $k$ to compensate for the weakness of viscous dissipation at these $k$.

\begin{figure}                                                                    
  \centerline{\includegraphics[width=0.50\textwidth]{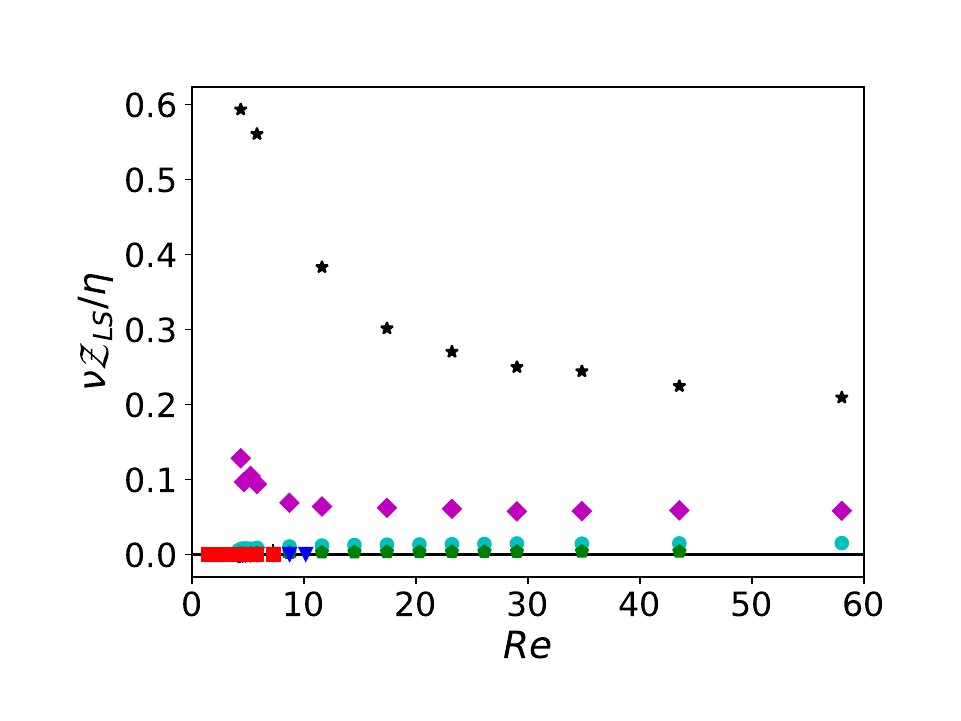}
              \includegraphics[width=0.50\textwidth]{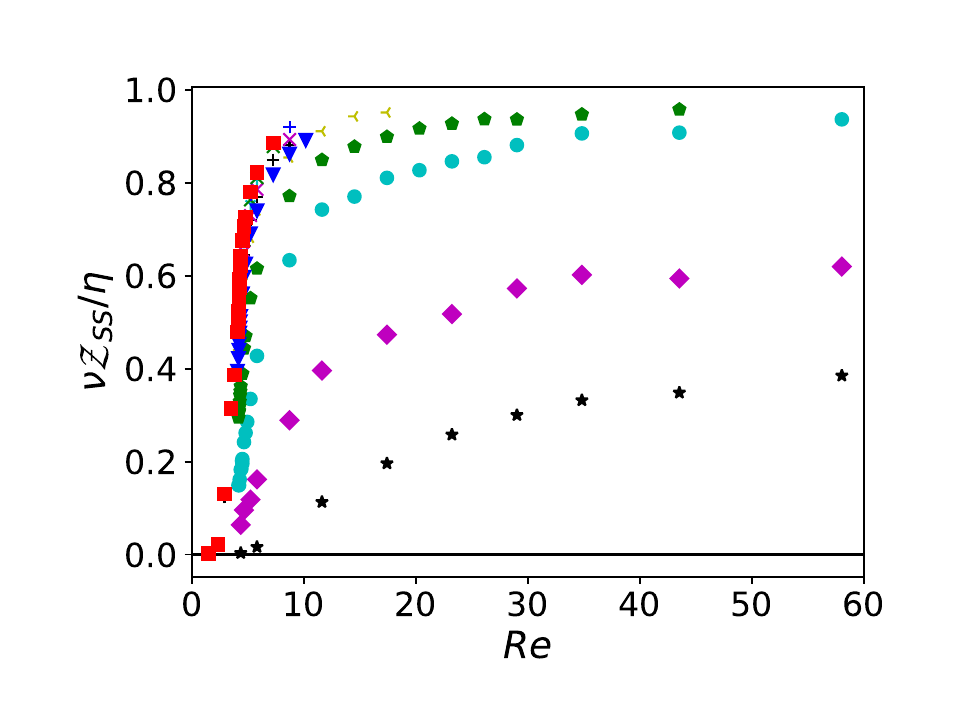} }
  \caption{Large-scale and small-scale palinstrophy bifurcation diagrams as a function of $Re$ normalized by $\eta/\nu$. Left: Large-scale palinstrophy $Z_{LS}$ defined as the palinstrophy in modes with $|k|\le k_{min}+3/L$. Right: Small-scale palinstrophy $Z_{SS}$ defined as the palinstrophy in modes with $|k|\ge k_f+6/L$.}
\label{fig:Palinstrophy}
\end{figure}

In Fig.~\ref{fig:Palinstrophy} we plot, in analogy with Fig.~\ref{fig:Enstrophy}, the large- and small-scale
enstrophy dissipation given, respectively, by $\nu \mathcal{Z}_{LS}/\varepsilon$ and $\nu \mathcal{Z}_{SS}/\varepsilon$. The figure confirms that large-scale enstrophy dissipation vanishes as
$Re\to\infty$ and $\Lambda\to \infty$ (left panel) and that almost all enstrophy dissipation takes place in the smallest scales (right panel). However, at moderate scale separation $\Lambda$ the fraction of the enstrophy dissipated at large scales remains finite. {In particular, it is known that the case $\Lambda=1$ where all energy and enstrophy injection occurs at the largest scale $|{\bf k}|=k_{min}$
is absolutely stable and that in this case all energy and enstrophy are dissipated at $k_{min}$ \cite{marchioro1986example}.}   
The general picture from these results is therefore that in the $Re\to\infty$, $\Lambda\to\infty$ limit
most of the enstrophy is dissipated in small scales while most of the energy
is dissipated at large scales, reaching amplitudes $\mathcal{E}\propto \varepsilon/\nu k_{min}^2 $. If the $\Lambda\to\infty$ limit is not taken, the fraction of the enstrophy dissipated at large scales remains finite.
%

\subsection{Spectra and Fluxes} 

\begin{figure}
  \centerline{\includegraphics[width=0.50\textwidth]{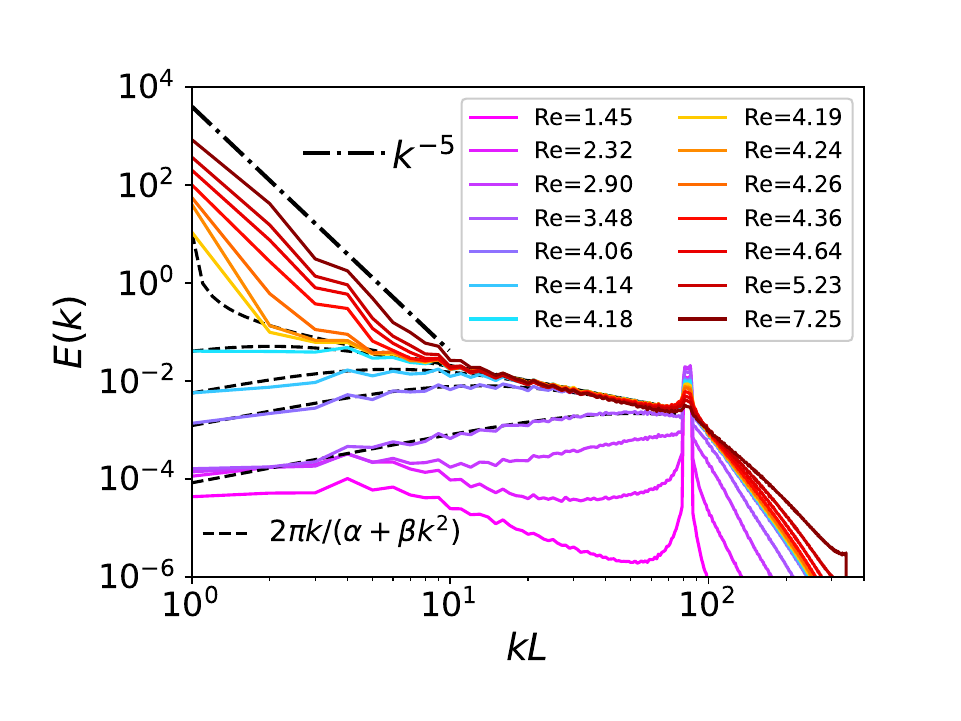}
              \includegraphics[width=0.50\textwidth]{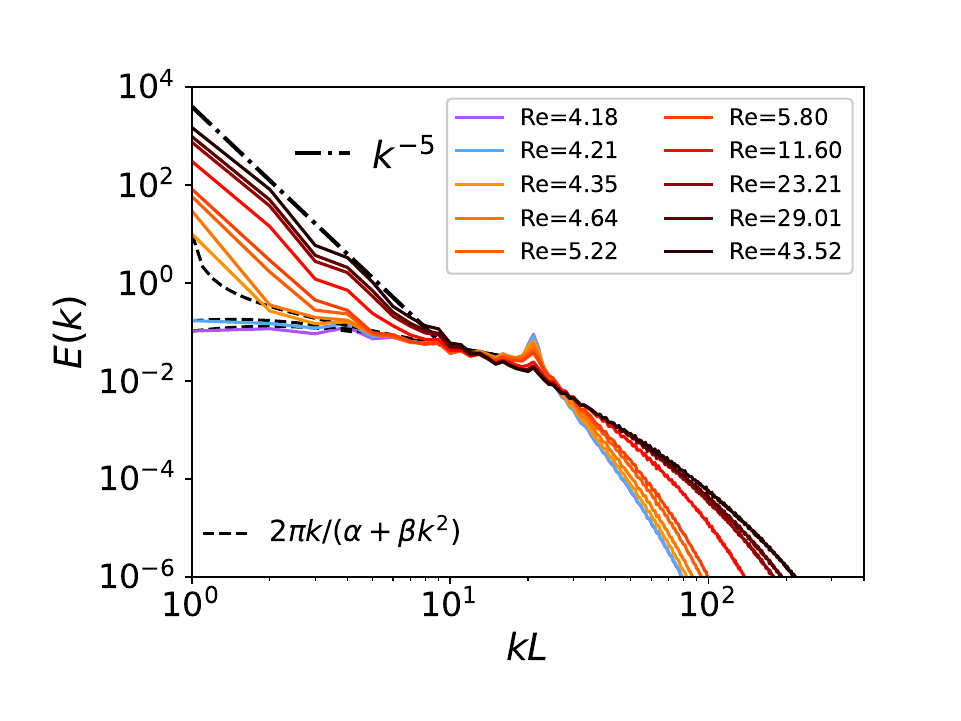} }
 \caption{Energy spectra for $k_{min}=1$ runs with $\Lambda=80$ (left) and $\Lambda=20$ (right). Dashed-dotted lines indicate the $k^{-5}$ power law. Dashed lines represent the best fit to the equilibrium spectra $E(k)=Ak/(k^2-k_s^2)$, where $A$, $k_s$ are fitting parameters and $k_s$ can be imaginary (i.e., $k_s^2<0$). 
\label{fig:Energy_Spec}}
\end{figure}

To better understand the observed behaviour we
plot in Fig.~\ref{fig:Energy_Spec} the energy spectra for different values of $Re$ and two values of the scale separation:
$\Lambda\equiv k_f L=80$ (left panel) which is the largest scale separation studied, and $\Lambda\equiv k_f L=20$ (right panel) for which sufficiently large values of $Re$ were reached to reveal the asymptotic behaviour of the energy (demonstrated in Fig.~\ref{fig:Energy}).
Different stages in the form of the energy spectrum are observed as the Reynolds number varies.
First, for very small $Re$, $Re \ll Re_c$, the dissipation dominates at all scales and the energy is concentrated at the
forcing scale with a fast exponential drop at larger wave numbers and a distribution at small wave numbers that is due to a "beating" effect from the forced modes. 
As $Re$ increases, approaching $Re_c$ from below,
the dissipation at large scales remains negligible, while remaining significant for the forced modes.  
The spectrum in this regime attains a form that is very close to Kraichnan's absolute equilibrium prediction, cf. Eq.~(\ref{eq:kraichnan_spectrum2D}). 
This behaviour continues even after the threshold $Re=Re_c$ is crossed,
leading to negative values of $\alpha/\beta$, where $\alpha$ and $\beta$ are the generalised temperature parameters obtained by fitting Kraichnan's spectrum to the DNS data. This observation provides support for the claim that the large-scale dynamics in two-dimensional
turbulence in the absence of large-scale friction can be described by equilibrium statistical physics for $Re$ close to $Re_c$.

\begin{figure}
  \centerline{\includegraphics[width=0.50\textwidth]{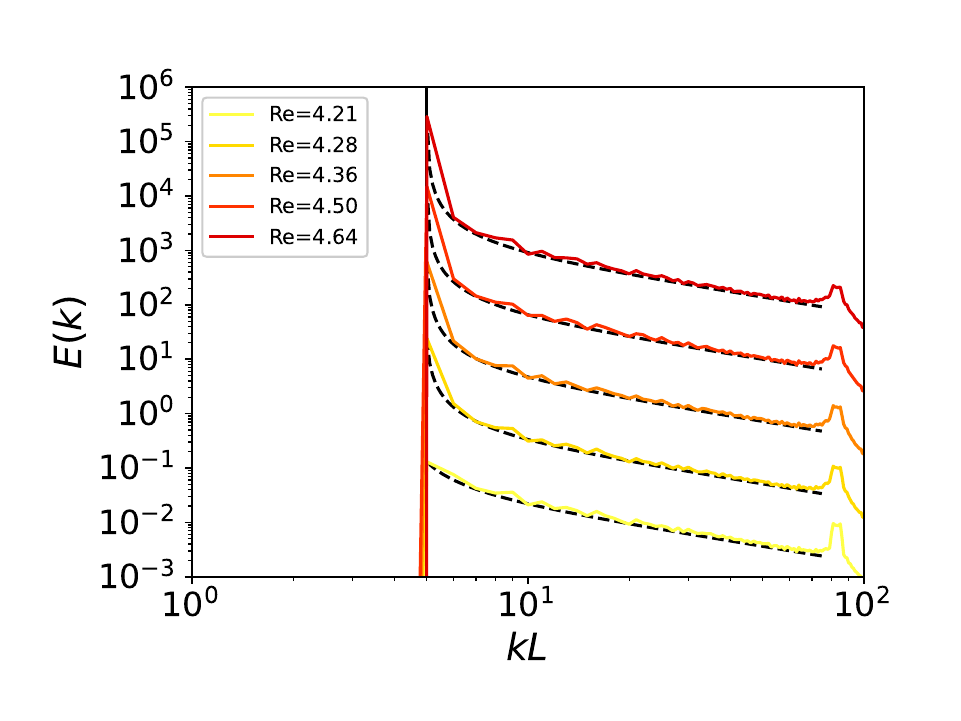}
              \includegraphics[width=0.50\textwidth]{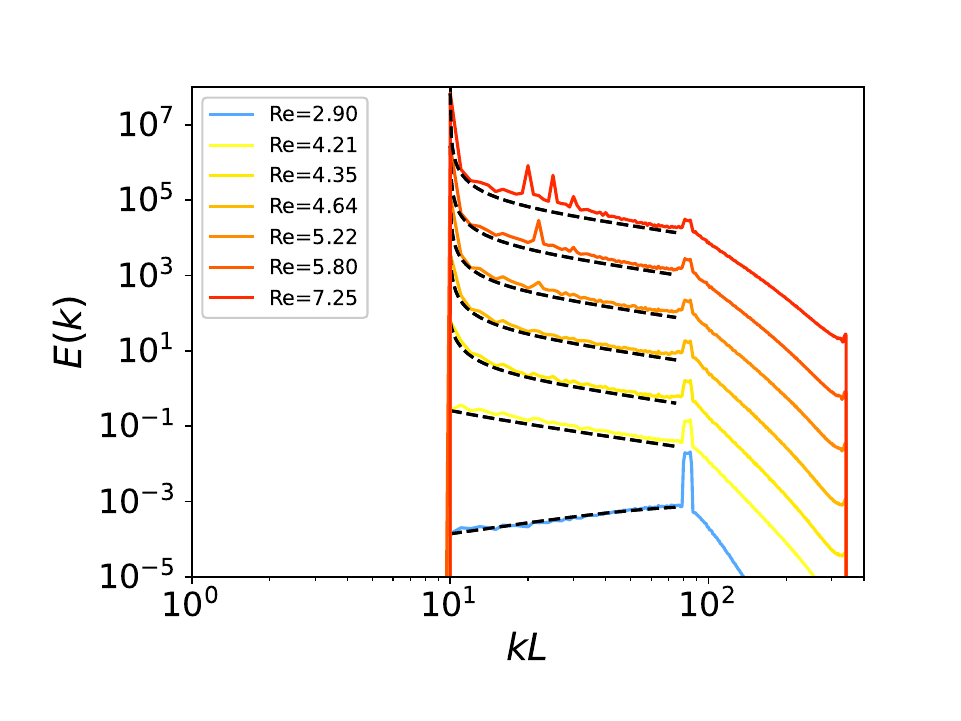} }
 \caption{Energy spectra for $\Lambda\equiv k_fL=80$ runs with $k_{min}L=5$ (left panel) and $k_{min}L=10$ (right panel). Dashed lines indicate the best fit to the equilibrium spectra $E(k)=Ak/(k^2-k_s^2)$. Spectra have been shifted vertically for clarity.
\label{fig:Energy_Spec_EQ}}
\end{figure}

\begin{figure}
  \centerline{\includegraphics[width=0.50\textwidth]{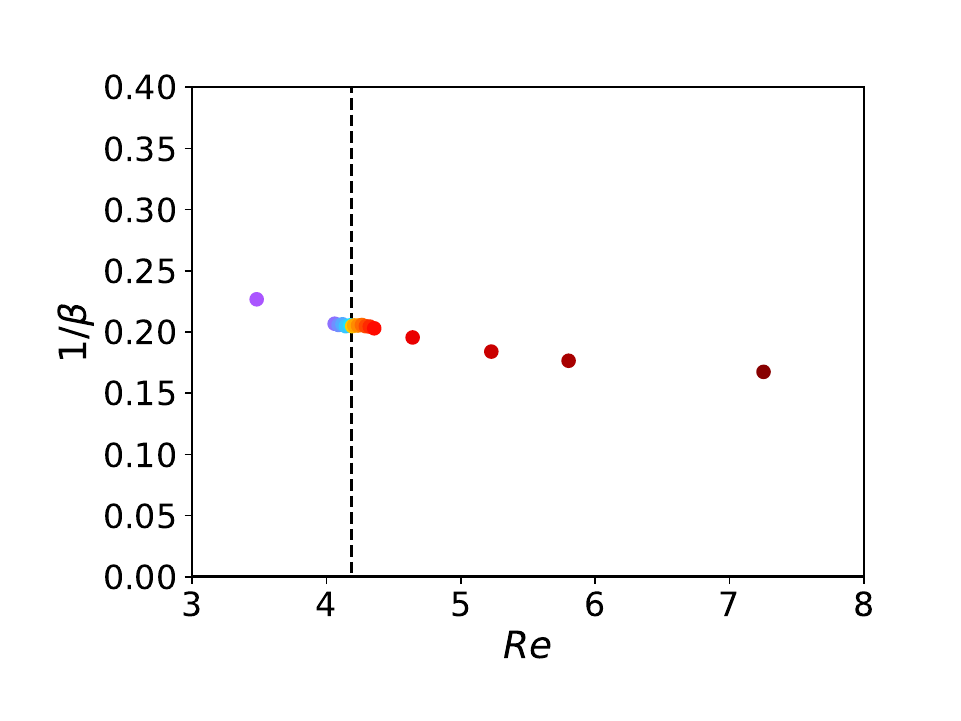}
              \includegraphics[width=0.50\textwidth]{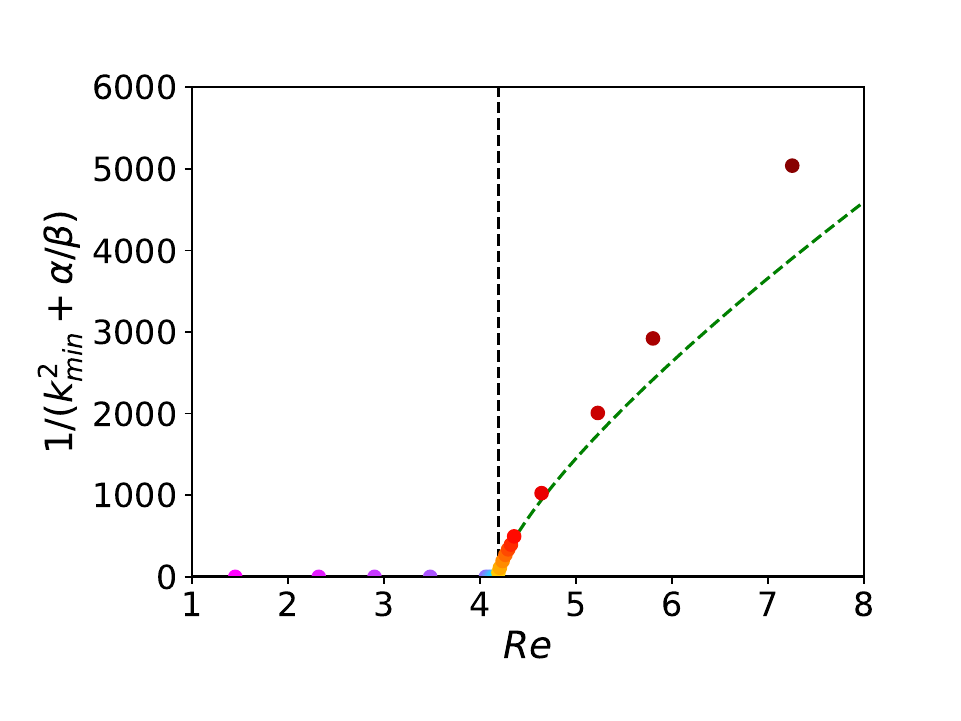}
}
 \caption{ 
Parameters of the fitted Kraichnan spectra as a function of $Re$ for $k_{min}L=1$, $\Lambda\equiv k_f L=80$.  Left panel: $1/\beta$ measuring the amplitude of the spectrum at large $k$ remains nearly unchanged as $Re$ crosses $Re_c$. Right panel: $1/(k_{min}^2+\alpha/\beta)$ increases sharply at $Re=Re_c$, indicating that the onset of the large-scale condensate is associated with a change in the location of the singularity in the denominator of Kraichnan's prediction.
 The colours of the data points are the same as those used in Fig.~\ref{fig:Energy_Spec_EQ}.}
\label{fig:Energy_Spec_KS}
\end{figure}

The Kraichnan solutions with negative $\alpha/\beta$
display a large peak in the spectrum at the minimum wave number. Flows with $k_{min}L=1$ have 
well-separated discrete wave numbers, which makes it hard to precisely validate the predicted functional form of the spectrum.
For this reason we plot in Fig.~\ref{fig:Energy_Spec_EQ} the cases $\Lambda\equiv k_fL=80$ with $k_{min}L=5$ and $k_{min}L=10$. Here the discreteness of the spectrum is less pronounced and it can be seen better how the Kraichnan solutions fit to the resulting spectrum for $Re$ close to $Re_c$. 
In Fig.~\ref{fig:Energy_Spec_KS}, we show how the parameters of Kraichnan's spectrum fitted to the DNS results at large scales change with $Re$. 
{The figure shows separately the overall spectral amplitude $1/\beta$ (left panel) and the quantity $(k_{min}^2+\alpha/\beta)^{-1}$ that expresses how close the singularity at $k^2=-\alpha/\beta$ is to $k_{min}^2$ (right panel).}
The figure shows that the amplitude $1/\beta$ changes little as $Re$ crosses $Re_c$, while the quantity $1/(k_{min}^2+\alpha/\beta)$ changes from close to zero below $Re_c$ to finite values above $Re_c$. This indicates that the onset of large-scale condensation at $Re=Re_c$ is associated with a shift in the location of the singularity in the denominator of the spectrum (\ref{eq:kraichnan_spectrum2D}), rather than an overall change in its amplitude.

As $Re$ is further increased and the condensate becomes stronger, the functional shape of the spectrum changes.
A new $k^{-5}$ power law appears that extends from the smallest wave number up to an intermediate wave number
that becomes larger and larger as $Re$ increases. In the right
panel of Fig.~\ref{fig:Energy_Spec}, where $\Lambda=20$ and larger values of $Re$ are reached, the $k^{-5}$ power law extends close to the forcing scale. As we show in Appendix~\ref{sec:app} and discuss further below in Sec.~\ref{sec:rad_profile}, this steep $k^{-5}$ power law at large scales, which is reported here for the first time (as far as we know), is due to the formation of a condensate with
a logarithmic vorticity profile that was recently predicted in \cite{doludenko2021coherent}. 
We remark that since energy is primarily dissipated at large scales $\nu \int_{k_{min}}^{k_f} E(k) k^2 dk \approx \varepsilon$, and hence that a spectrum of the form $E(k)=Ak^{-5}$ implies that the prefactor $A$ at large $Re$ becomes $A=2\varepsilon k_{min}^2/\nu $.
Furthermore the fraction of enstrophy dissipation occurring at large scales, $\nu\int_{k_{min}}^{k_f} E(k) k^4 dk/\varepsilon k_f^2$, becomes proportional to $\Lambda^{-2} \ln(\Lambda)$, in agreement with our observation that enstrophy is dissipated primarily at small scales only when the scale separation factor $\Lambda\equiv k_f/k_{min}$ is sufficiently large.
We emphasise finally that the $k^{-5}$ profile depends on the form of the dissipation. If a weak bottom drag is considered instead, the vorticity profile becomes proportional to $1/r$ and the resulting spectrum in the statistically stationary state is $k^{-3}$, as  confirmed numerically, see e.g. \cite{parfenyev2024statistical}. The sensitivity to the type of dissipation reflects the fact that the condensate at high $Re$ in this driven-dissipative setup is not a quasi-equilibrium state: as $Re$ increases, the system transitions from a quasi-equilibrium state to an out-of-equilibrium state.

\begin{figure}
  \centerline{\includegraphics[width=0.50\textwidth]{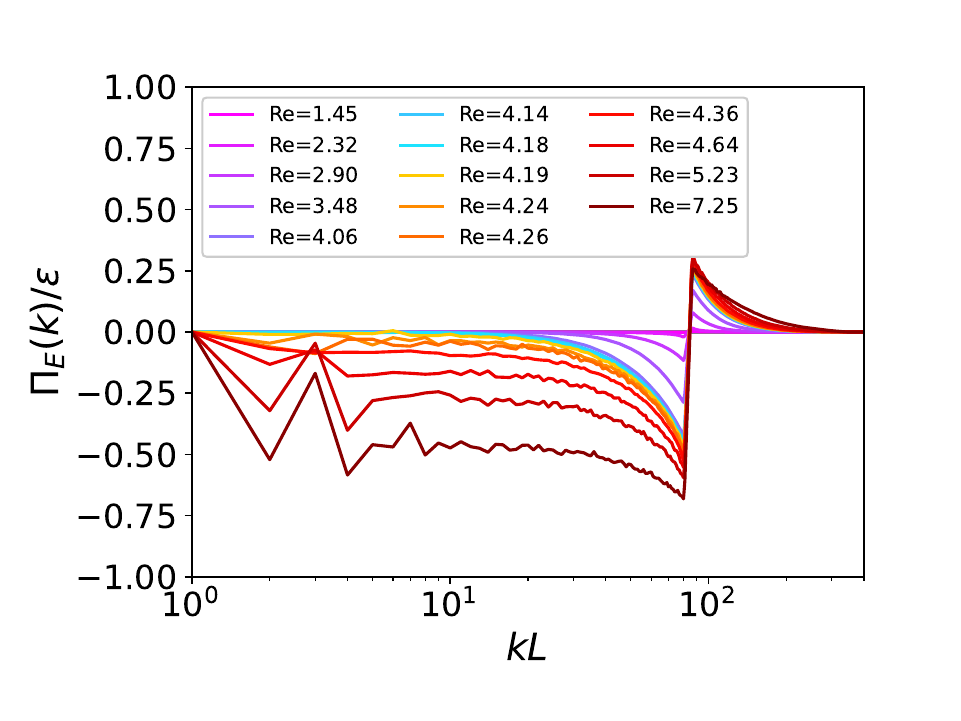}
              \includegraphics[width=0.50\textwidth]{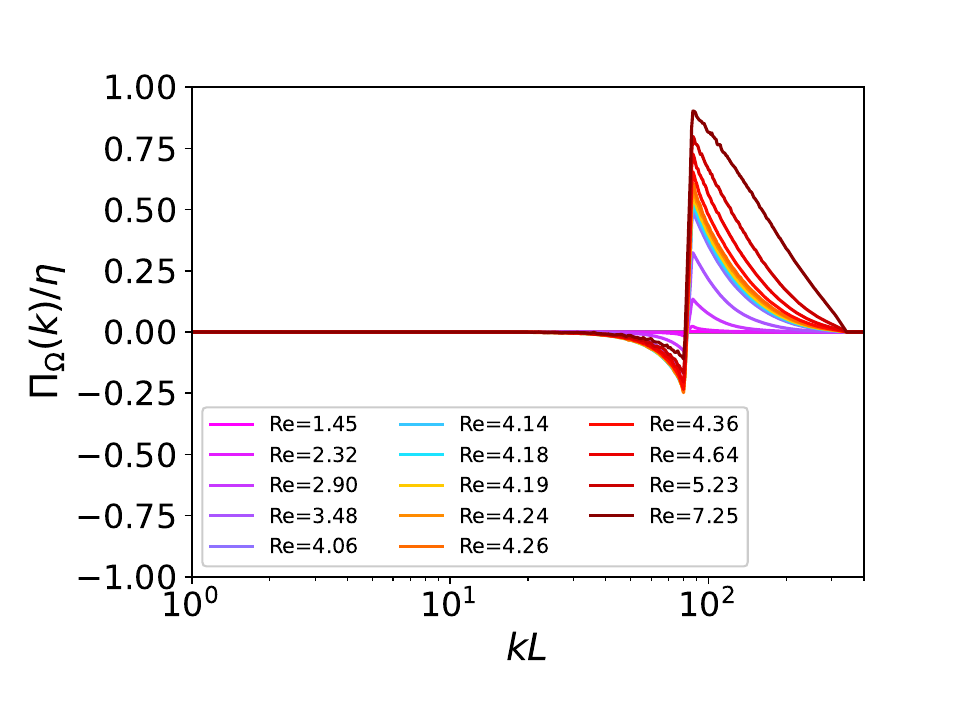} }
  \centerline{\includegraphics[width=0.50\textwidth]{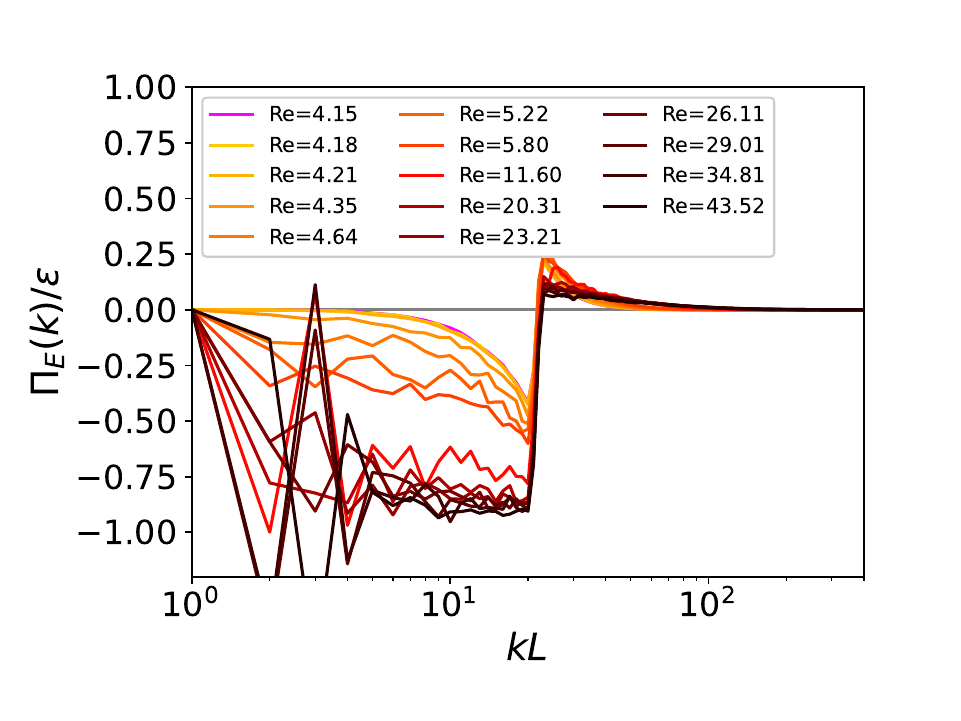}
              \includegraphics[width=0.50\textwidth]{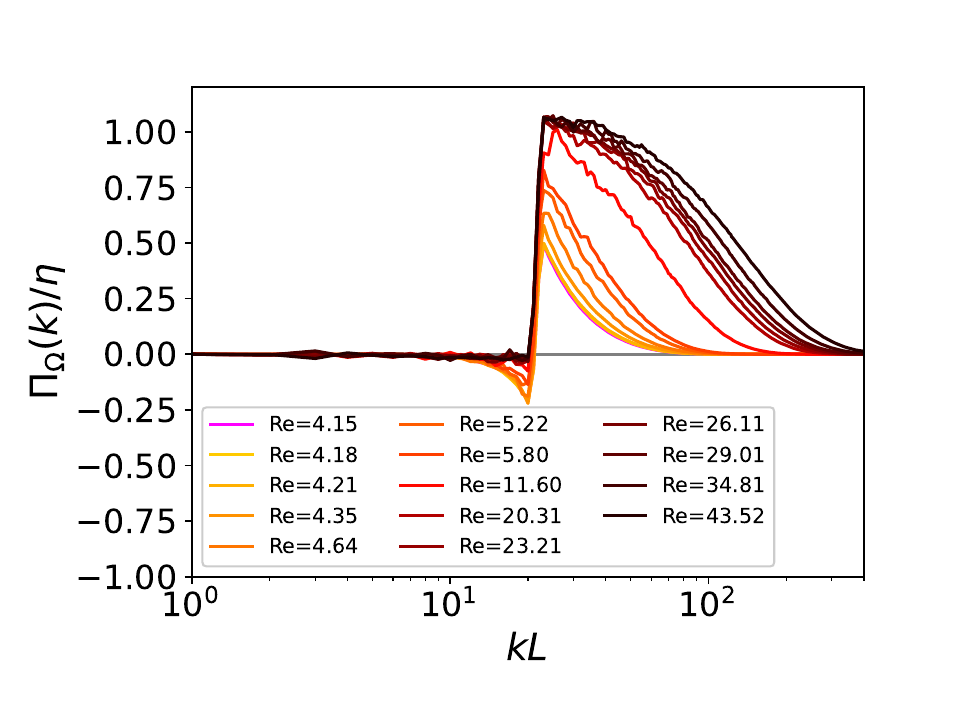} }
 \caption{Energy (left) and enstrophy (right) fluxes for $\Lambda\equiv k_f L=80$ (top row) and $\Lambda\equiv k_f L=20$ (bottom row) for different $Re$.
 }
\label{fig:Energy_Spec_KS2}
\end{figure}

The out-of-equilibrium nature of this system is associated with finite mean fluxes of energy and enstrophy across scales, shown in Fig.~\ref{fig:Energy_Spec_KS2}, averaged over the stationary state. The energy flux due to nonlinear interactions is given by $\Pi_E(K) = -\langle \mathbf{u}^{<K}\cdot(\mathbf{u}\cdot\nabla\mathbf{u})\rangle $, where $\mathbf{u}^{<K}$ has been low-pass filtered to contain only Fourier modes with wave numbers of modulus less than $K$. The enstrophy flux is defined similarly as $\Pi_\Omega(K)=-\langle \omega^{<K} (\mathbf{u}\cdot \nabla \omega)\rangle $, where $\omega^{<K}$ is the low-pass-filtered vorticity. The energy flux $\Pi_E$ becomes negative at scales larger than the forcing scale ($\Lambda\equiv k_f L =80$ in the top row, $\Lambda\equiv k_f L=20$ in the bottom row) as $Re$ increases beyond $Re_c$, indicating an inverse (i.e., upscale) energy transfer at a rate that increases with $Re$. There is a large variance in the energy flux at large $Re$, which leads to fluctuations that remain visible at low $k$ and could only be further reduced by extensive averaging. The magnitude of the upscale flux approaches the energy injection rate $\varepsilon$ as $Re$ increases. The enstrophy, on the other hand, is transferred towards smaller scales at a rate that also increases with increasing $Re$. Here, the variance is smaller and the enstrophy flux approaches the enstrophy injection rate $\eta$ at large $Re$.   
{We note that the constant inverse flux of energy and the constant forward flux of enstrophy demonstrated here is far from a trivial
result in a system where no large scale dissipation mechanism is present and the system relies on viscous dissipation to saturate the large-scale energy growth. Two-dimensional turbulence only achieves this state by reaching sufficiently large condensate amplitude with sufficiently steep spectra so that all the energy injected is dissipated in just a few small wave number modes.}

\subsection{Radial vorticity Profile} 
\label{sec:rad_profile}
The spectral viewpoint of the condensate state described above is complementary to a physical-space picture of the system. Recent investigations of the forced-dissipative condensate in the context of quasi-linear theory have led to explicit expressions for the spatial profile of the condensate vortices, with differing predictions depending on the dissipation mechanism saturating the growth of the condensate, as discussed earlier. In the present case of a viscously saturated condensate without bottom drag, Doludenko et al.~\cite{doludenko2021coherent} predict a vorticity profile in the condensate state of the form 
\begin{equation} 
\overline{\omega}(r) =  - 2 \Sigma ( \log(r/L)+c_0),
\label{eq:log}
\end{equation} 
where $\overline{\omega}(r)$ is the time-averaged vorticity 
{at distance $r$ from the center of the condensate.}
The constant $\Sigma$ has the dimension of vorticity and in the $Re\to \infty$, $\Lambda \to \infty$ limit asymptotes to $\Sigma = \Sigma^*\equiv  \pm \sqrt{\frac{\varepsilon}{\nu}}$, where the sign differs between the two counter-rotating large-scale vortices. Finally, $c_0$ is an order one constant. Expression (\ref{eq:log}) is valid for $1\ll k_fr \ll k_f L$.

\begin{figure}
  \centerline{\includegraphics[width=0.50\textwidth]{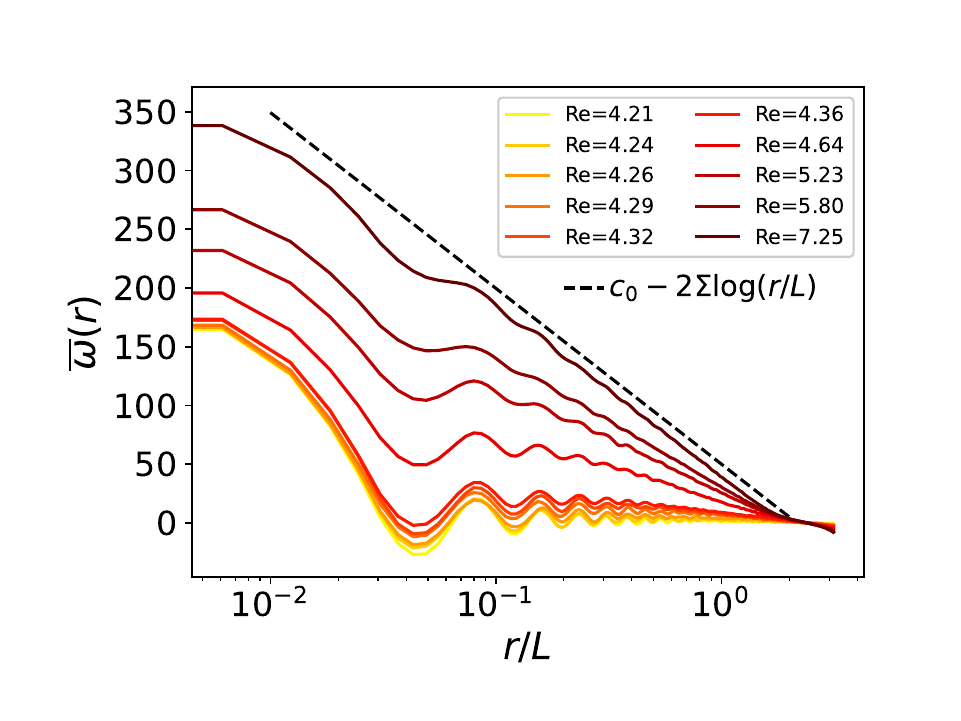}
              \includegraphics[width=0.50\textwidth]{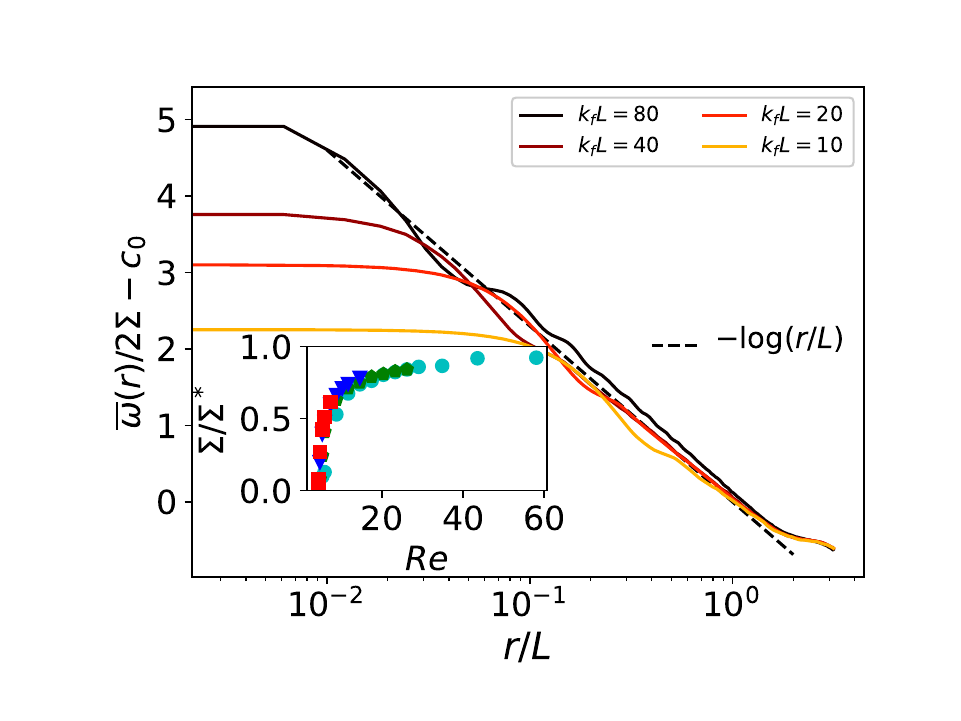} }           
 \caption{Left panel: Vorticity profile $\overline{\omega}(r)$ of the condensate for different $Re$ (left) and $k_fL=80$. The dashed line indicates the predicted functional form.
 Right panel: $\overline{\omega}(r)$ normalised by the measured quantity $2\Sigma$ for different values of $\Lambda=k_f L$ (right) at the maximum value of $Re$ reached. The inset shows the measured value of $\Sigma$ normalised by $\Sigma^*\equiv\sqrt{\varepsilon/\nu}$.
\label{fig:Vprofile}}
\end{figure}

In Fig.~\ref{fig:Vprofile} we plot the mean vorticity profile $\overline{\omega}(r)$ from different runs. 
To measure $\overline{\omega}(r)$ we trace for each instant of time the maximum and minimum value of the vorticity, and measure the mean vorticity at all points in the grid at distance $r$ from these maxima. 
This procedure was repeated many times in the stationary state, and the time-averaged vorticity profile was calculated.
The left panel displays the vorticity profile from the runs with the largest scale separation $k_fL=80$.
The right panel shows $\overline{\omega}(r)$ normalised by the measured $2\Sigma$ for different values of $\Lambda=k_f L$ (right panel) and the maximum value of $Re$ attained at these $\Lambda$.
For small $Re$ spatial oscillations at the forcing scale are observed. As $Re$ increases the amplitude around $r=0$ increases and for sufficiently large $Re>Re_c$ the predicted functional form of $\overline{\omega}(r)$ is confirmed (left panel) and this is so over an increasingly wide range of radii at large $Re$ as the scale separation factor $\Lambda$ between the forcing scale and the domain size increases (right panel), in agreement with the prediction of Doludenko et al. \cite{doludenko2021coherent}. The inset shows the measured value of the slope $\Sigma$ normalised by $\Sigma^*$ demonstrating that $\Sigma \approx \Sigma^*$ at large $Re$.
 
The spatial structure and the spectral properties of the condensate are naturally linked via the Fourier transform. In Appendix~\ref{sec:app}, we directly compute the Fourier transform of the logarithmic vorticity profile and show that this leads to an energy spectrum with a $-5$ exponent in agreement with the large-scale form of the energy spectra obtained in our DNS; see Fig.~\ref{fig:Energy_Spec}. In contrast, when the condensate is saturated by a linear bottom drag, in which case the vorticity profile is of the form $\omega(r)\propto 1/r$, a $k^{-3}$ spectrum is found instead, in agreement with the literature; see, e.g., \cite{parfenyev2024statistical}.
It is interesting to note that the $-3$ exponent coincides with the value reported by Chertkov et al. for the case with no bottom drag or any other dissipation mechanism active at large scales in their hyperviscous simulations \cite{chertkov2007dynamics}. We surmise that these simulations were likely limited to the transient regime and failed to reach the final statistically stationary state. The agreement between the spectral exponents reported in the transient regime and the saturated condensate with bottom drag is thus coincidental since, as discussed above, the energy spectrum in the stationary state depends explicitly on the dissipation mechanism.

%
\section{Conclusions}
\label{sec:conclusions}
In this article we have characterised the viscously saturated condensate in stochastically forced two-dimensional turbulence using extensive direct numerical simulations while varying independently both the 
Reynolds number $Re$ and the scale separation $\Lambda$ between the forcing scale and the domain size. The purpose of this work is to determine whether the condensate dynamics can be approximated by inviscid equilibrium theory
or by quasi-linear theory wherein there is a local balance between energy injection and viscous dissipation.
Our results establish that for sufficiently large scale separations, one or other of these two theories applies but over different ranges of $Re$.

When $Re$ is close to its critical value for the onset of turbulence, large scales are well described by equilibrium dynamics.
This is true both below and above onset. Above onset, a condensate forms but is predominantly composed of single-scale Fourier modes whose amplitude increases with $Re$.
We emphasise that while the forcing-scale Reynolds number $Re$ is of order one in this range, the effect of viscosity on the large scales
is still negligible and so these scales are well described by inviscid dynamics. This becomes more and more so as the scale separation $\Lambda$ increases.
Close to the onset of turbulence at the critical Reynolds number $Re=Re_c\approx 4.19$, we find a self-similar scaling of the kinetic energy with $Re-Re_c$ with an exponent smaller than unity. The energy spectrum on large scales follows Kraichnan's absolute equilibrium prediction for $Re$ close to $Re_c$. 

As $Re$ increases further and the condensate becomes more dominant, the energy spectrum deviates from this equilibrium form, showing a steep power-law range at low wave numbers with exponent $-5$.
In this range of $Re$, most of the energy dissipation occurs within the condensate and so takes place at large scales supplied by a constant energy flux from the forcing scale to the largest scales of the system. In physical space, the condensate assumes a logarithmic profile in agreement with the quasi-linear theory prediction \cite{doludenko2021coherent}. 
We have shown that the spectral exponent $-5$ is a direct consequence of this logarithmic radial vorticity profile 
while the exponent $-3$ is found instead when the growth of the condensate is arrested by weak bottom drag. 
The deviations from the equilibrium predictions as well as the dependence of the condensate functional form 
(in real and Fourier space) on the dissipation mechanism indicate that in the large $Re$ limit, 
the large scales of forced-dissipative two-dimensional turbulence are out-of-equilibrium and are better described by quasi-linear theory.

Several open questions for these viscously saturated condensates remain.
The critical scaling exponent of the kinetic energy near the onset of two-dimensional turbulence was only determined numerically and remains to be explained on theoretical grounds. Although this work focused on the case of two dimensions, the dependence of the results on the inclusion of a third dimension remains to be investigated, both in a quasi-two-dimensional setting \cite{alexakis2018cascades,alexakis2023quasi,van2024phase}, for instance, in a thin-layer geometry \cite{van2019condensates,musacchio2019condensate} or under the influence of rapid rotation \cite{seshasayanan2018condensates}, even though such studies will require significantly greater computational resources. Another important future direction is to examine the robustness of the results presented here for other forcing mechanisms and in the presence of linear or nonlinear bottom drag, e.g., in the context of instability-driven turbulence \cite{van2022spontaneous,van2023vortex}, including turbulent convection \cite{guervilly2014large,rubio2014upscale,favier2014inverse}, and active turbulence \cite{alert2022active}, where a condensate can also form from an inverse cascade, depending on parameters \cite{linkmann2019phase,linkmann2020condensate,puggioni2022giant}.




\ack{This work was granted access to the HPC resources of GENCI-TGCC \& GENCI-CINES (Project No. A0150506421), and the Purdue Anvil CPU cluster \cite{song2022anvil} as part of NSF ACCESS (project number PHY230056). This work has also been supported by the Agence nationale de la recherche (ANR DYSTURB project No. ANR-17-CE30-0004), by the National Science Foundation (Grants DMS-2009563 and DMS-2308337 (EK)), by the German Research Foundation (DFG Projektnummer: 522026592) and the France-Berkeley Fund (Project \#8 2023).}

\subsubsection*{Declaration of Interests}
The authors report no conflict of interest.

\appendix 
\section{Computation of energy spectrum from spatial profiles}
\label{sec:app}
\subsection{The case of a viscously saturated condensate without bottom drag}
Doludenko et al. \cite{doludenko2021coherent} 
state that the viscously saturated condensate vortex has the following azimuthal velocity profile as a function of radius $r$
\begin{equation}
    U(r) = -\Sigma r \ln(R/r), \qquad \text{provided } k_f^{-1} \ll r \ll R, \label{eq:U_large_r}
\end{equation}
with $\Sigma \propto \sqrt{\varepsilon/\nu}$, provided  where $k_f L\gg 1$ is the forcing wave number, $R$ is comparable to the domain size with $R/L\sim O(1)$
and $Re\gg1$. The coefficient $\Sigma$ is positive in the cyclone and negative in the anticyclone. For simplicity, we compute the Fourier transform of vorticity and consider the enstrophy spectrum, which is related to the energy spectrum by a factor of $k^2$.\\
In polar coordinates, for an azimuthal velocity $\mathbf{u} = u(r)\hat{e}_\theta$, the vorticity is given by $\omega = \frac{1}{r} \frac{\partial}{\partial r} (r u(r))$
so for $k_f^{-1}\ll r \ll R$ the vorticity is given by
\begin{align}
    \omega(r) &= -\Sigma (2\ln(L/r)+c_0), \label{eq:vort_radial_profile} 
\end{align}
where the arbitrariness of $R$ has been absorbed into the coefficient $c_0$.
In the small region $k_fr\sim O(1)$ the vorticity field is approximately constant with $\omega(r) = \omega_0\equiv-2\Sigma \ln(k_fL)$.
For $k$ in the range $1/L\ll |{\bf k}| \ll k_f$
contributions from $r=\mathcal{O}(L)$ are subdominant and neglected. A radially symmetric function has a radially symmetric Fourier transform. Therefore, we seek to compute
\begin{equation}
   \hat{\omega}(k) = \frac{1}{2\pi}\int_0^{2\pi} d\theta \int_0^L r\, dr\,  \omega(r) e^{ikr \cos\theta}  = \int_0^L rJ_0(kr) \omega(r) dr.
\end{equation}
%
Performing the integration gives the Fourier transform
\begin{equation}
\hat{\omega}(k)= 
  \frac{2\Sigma}{k^2} J_0(k/k_f) + o(J_0(kL)).
\end{equation}
 For $1/L \ll k \ll k_f$ 
 (since $J_0(0) = 1$ and $J_0(x)\stackrel{x\to \infty}{\longrightarrow} 0$),
 the $o(J_0(kL))$ terms can be neglected and this
 leads to a spectral exponent of $\Omega(k) = |\hat\omega(k)|^2\propto k^{-4} $, implying an energy spectrum that scales with $k$ as
 \begin{equation}
     E(k) \propto k \frac{\Omega(k)}{k^2} \propto k^{-5}.
 \end{equation}
This is consistent with the findings of the DNS reported in the main text. We  also have synthetically generated a logarithmic vorticity profile and numerically computed the associated energy spectrum to confirm this result.

 \subsection{The case of bottom drag}
 For the case with bottom drag (also known as Rayleigh damping), Laurie et al. \cite{laurie2014universal} showed that for $k_f^{-1} \ll r \ll R$, where $R$ is again comparable to the domain size $L$, the azimuthal velocity is approximately constant $U(r) = \sqrt{3\varepsilon/\alpha}$, i.e. vorticity $\omega(r) = \frac{1}{r} \partial_r (r U(r)) = \sqrt{3\varepsilon/\alpha}/r$, where $\varepsilon$ is the energy injection rate and $\alpha$ is the Rayleigh damping coefficient. 
 The resulting Fourier transform of the vorticity reads
\begin{align}
     \hat{\omega}(k) \approx & \int_{\ell_f}^R r \omega(r) J_0(kr) dr  \\
     =& \sqrt{\frac{3\varepsilon}{\alpha}} \frac{1}{k} \left[  {}_1F_2\left( \frac{1}{2}; 1,\frac{3}{2}; - \frac{y^2}{4}\right) y \right]_{y=k/k_f}^{y=kR}\\
     \approx & \sqrt{\frac{3\varepsilon}{\alpha}} R\quad  
 {}_1F_2 \left( \frac{1}{2}; 1,\frac{3}{2}; - \frac{k^2R^2}{4}\right). 
 \end{align}
Since the generalised hypergeometric function $ {}_1F_2 \left( \frac{1}{2}; 1,\frac{3}{2}; - \frac{x^2}{4}\right)=O(1/x)$ for $x\to \infty$  (see \cite{mathai2006generalized}), and $ {}_1F_2 \left( \frac{1}{2}; 1,\frac{3}{2}; - \frac{x^2}{4}\right)$ remains bounded as $x\to 0$, the energy spectrum in this case is found to scale as
\begin{equation}
    E(k)\sim k |\hat{\omega}(k)|^2/k^2 \sim k^{-3}.
\end{equation}
This scaling is less steep than in the case of viscous saturation, which is to be expected, since viscosity acts significantly only at the largest scales, allowing for a stronger accumulation of energy at low $k$. We have verified this conclusion by numerically computing the Fourier transform of a synthetic vorticity field with an $\omega(r)\propto 1/r$ range.




\enlargethispage{20pt}


\bibliographystyle{unsrt}
\bibliography{bibliography}

\end{document}